\documentclass[aps,prb,twocolumn,groupedaddress,showpacs]{revtex4-1}

\usepackage{CJK}
\usepackage{graphicx}
\usepackage{amsfonts}
\usepackage{amsmath}
\usepackage{amssymb}
\usepackage{bm}

\begin{document}
\begin{CJK}{GBK}{song}
\title{Long-Range triplet Josephson Current driven by the bias voltage}
\author{Hao Meng}
\author{Xiuqiang Wu}
\author{Feng Mei}
\affiliation{National Laboratory of Solid State Microstructures and
Department of Physics, Nanjing University, Nanjing 210093, China}
\date{\today }

  \begin{abstract}
  We study the long-range triplet Josephson current in a clean junction composed of two s-wave superconductors and a normal-metal/ferromagnet/normal-metal trilayer. Through applying the bias voltages on the metal regions by two antiparallel half-metal electrodes, we show that the amplitude and direction of this long-range current can be controlled easily and flexibly. Such current arises from the fact that the applied voltage can produce a nonequilibrium spin-dependent quasiparticle distribution in the metal regions so that the Cooper pairs acquire an extra momenta, which will lead to a spin-flip processes in the metal regions. This processes can produce the parallel spin triplet pairs in the central ferromagnet layer. In particular, if the voltage is applied only on one metal region, we further find that the recently discovered long-range superharmonic Josephson current will appear because of the transport of an even number of parallel spin triplet pairs.
  \end{abstract}

  \maketitle

  In recent years, the interplay between ferromagnetism and superconductivity in hybrid structures has been extensively studied because of the underlying rich physics and potential applications in spintronics~\cite{Buz,BerRMP}. In a homogeneous ferromagnet (F) adjacent to an s-wave superconductor (S), the Cooper pairs, consisting of two electrons with opposite spins and momenta, can penetrate into the F a short-range. In this case, the Cooper pair $\mid\uparrow\downarrow\rangle$ inside the F will acquire a total momentum $\emph{Q}$ as a response to the exchange splitting $2\emph{h}$ between the spin up and spin down bands~\cite{Esc}, where $Q\simeq2\emph{h}/\hbar{v_F}$, $v_F$ is the Fermi velocity. The resulting state is a mixture of singlet component and triplet component with zero total spin projection: $(\mid\uparrow\downarrow\rangle$$-$$\mid\downarrow\uparrow\rangle)\cos(Q$$\cdot$$R)$$+$$i$$\cdot$$(\mid\uparrow\downarrow\rangle$$+$$\mid\downarrow\uparrow\rangle)\sin(Q$$\cdot$$R)$. These two components oscillate in F with a same period but their phases differ by $\pi/2$.

  In contrast, the interested long-range triplet component with parallel electron spins can be induced by an inhomogeneous ferromagnetism, with the length scale approaching the coherence length of the normal-metal (N), which is typically about several hundred nanometers~\cite{Keizer,Robinson,Khaire,Klose}. Many different inhomogeneous configurations have been proposed recently for studying such current, including the F with a magnetic domain wall~\cite{BerRMP,BerPRL} or a spiral magnetic structure~\cite{Mohammad,Halasz}, the Josephson junction with multilayers of Fs~\cite{Hou,Volkov} or the spin active interface~\cite{Eschrig,Asano}. Specifically, for a Josephson junction with three ferromagnetic layers, if the direction of the magnetization in the interface F layer deviates from the one in the central F layer, a spin-flip scattering processes at the interface layers will arise due to such noncollinear magnetization. This process can convert the triplet pairs $\mid\uparrow\downarrow\rangle$$+$$\mid\downarrow\uparrow\rangle$ into the parallel spin pairs $\mid\uparrow\uparrow\rangle$ and $\mid\downarrow\downarrow\rangle$, which propagate coherently over long distances into the central F layer~\cite{Hou,Eschrig}. Indeed, many recent experiments have demonstrated this physical process and observed a strong enhancement of the long-range spin triplet supercurrents~\cite{Keizer,Robinson,Khaire,Klose,Anwar} based on an inhomogeneous ferromagnetism. It is also noted that a long-range Josephson current in S/F/S junction can be generated by the propagation of the opposite-spin Cooper pairs through injecting spin current into F~\cite{BobPRL}. However, this process requires an extreme experimental condition on the voltage so that the shifted value of the Fermi level for spin-up and spin-down subbands is equal to the exchange field of F.

  In this paper, we show that a long-range triplet Josephson current in a clean S/N/F/N/S junction can be generated and driven by the bias voltages, which are applied on N regions by two additional antiparallel half-metal (HM) electrodes. By tuning this applied bias voltages, we further show that the amplitude and direction of this Josephson current could be controlled easily and flexibly. This method can provide a very promising tool to artificially manipulate the Josephson current and is very important for the practical application of Josephson-current-based spintronics devices. The origin of this long-range current here is that the voltage in N regions can produce and maintain a nonequilibrium spin-dependent quasiparticle distribution. This distribution can provide the Cooper pair in N regions an additional momenta $Q$, where $Q$ is along the polarized direction of the HM electrodes and is perpendicular to the magnetization of the central F. The whole process will lead to the spin-flip scattering in N regions. This behavior can produce the parallel spin pairs with $S=\pm1$ in the central F layer and make the current become long-range. It is also found that the recently discovered superharmonic Josephson current~\cite{Tri,Richard} will appear when applying the voltage on only one N region. This is because of the phase coherent transport of an even number of parallel long-range spin triplet pairs.

\begin{figure}
  \centering
  \includegraphics[width=3.0in]{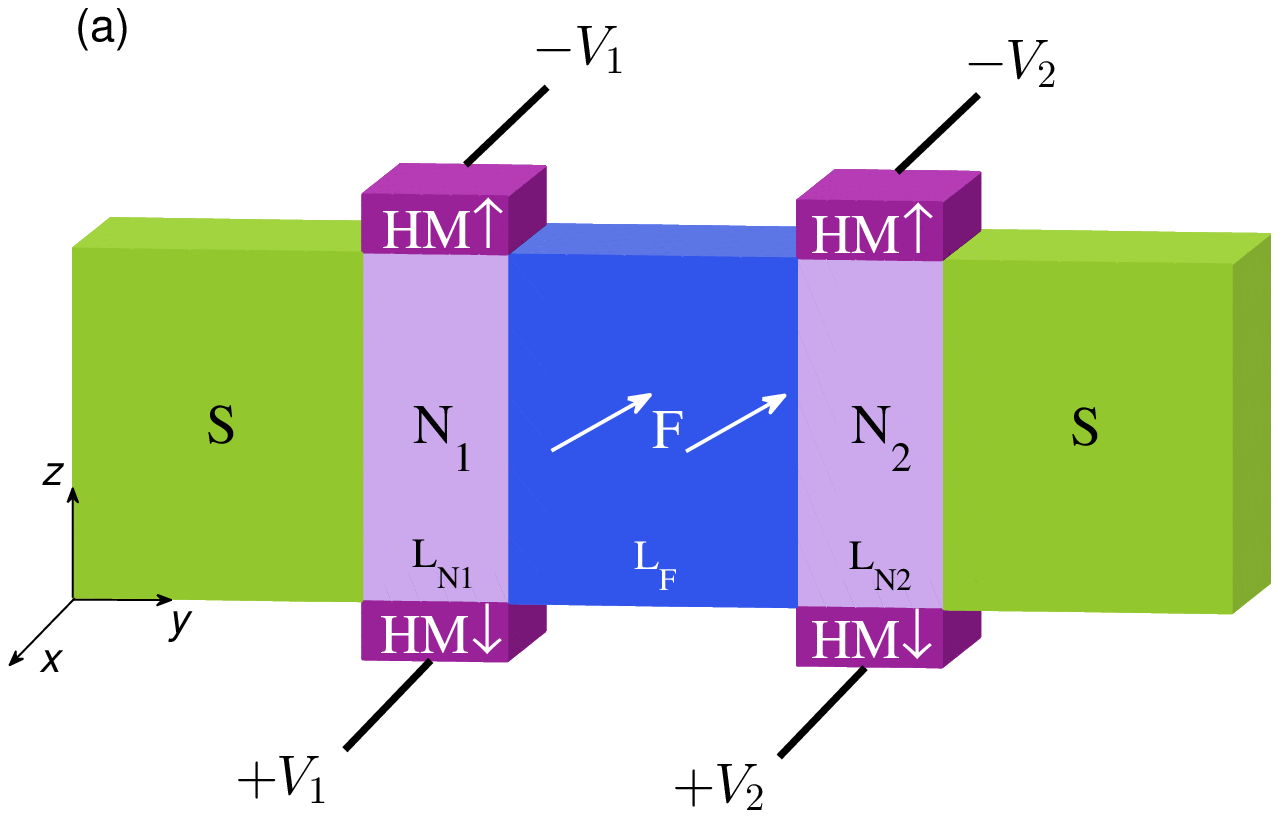} 
  \includegraphics[width=2.6in]{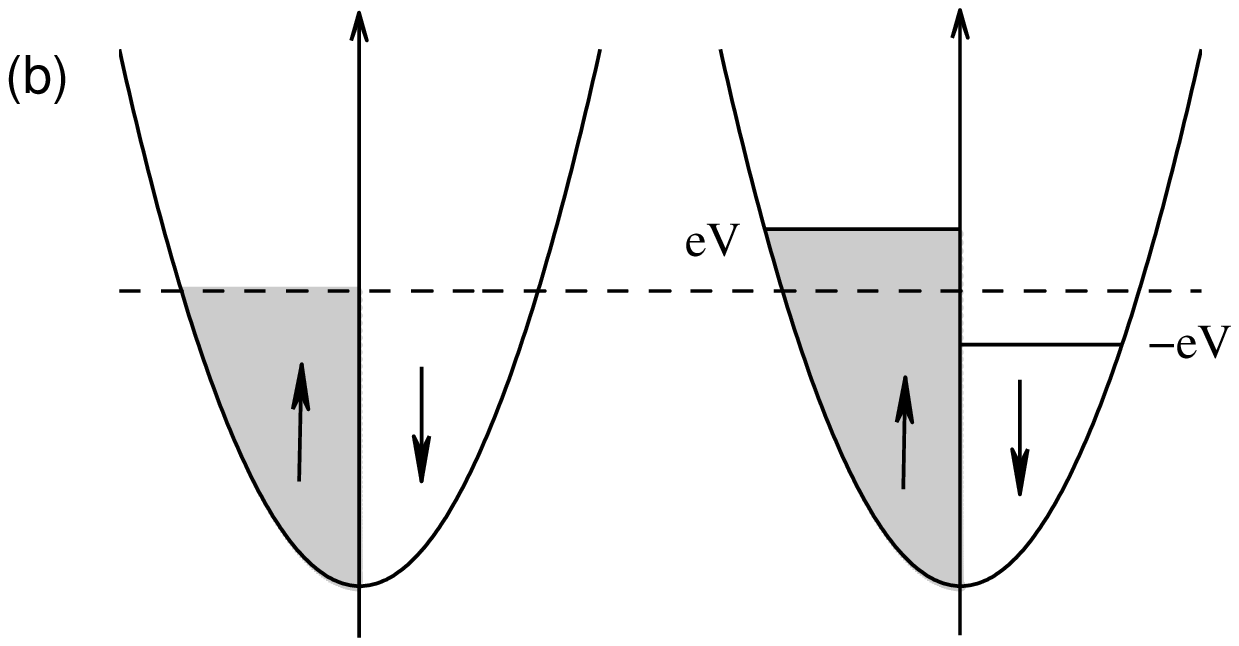} 
  \caption{(color online) (a) Schematic diagram of the S/N/F/N/S junction. The bias voltages are applied on the left and right N regions by two antiparallel HM electrodes  The phase difference between the two Ss is $\phi=\phi_L-\phi_R$. (b) The spin-up and spin-down quasiparticle distribution in N region without applying (left) and with applying (right) the bias voltage $V$.}
  \label{fig.1}
  \end{figure}

  In what follows, we assume that the HMs are polarized along the $z$ axis and their magnetization for up and down electrodes are antiparallel, whereas the F is oriented along the $-\emph{x}$ axis (see fig.~\ref{fig.1}(a)). When the polarizations of the HM electrodes are perpendicular to the ferromagnetic magnetization axis, Mal'shukov and Brataas~\cite{Mal} have show that the dissipative current does not penetrate deeply into the F. Furthermore, the dissipative current in our system is proportional to the spin current carried by spin polarized electrons through the F. Recent experiment has demonstrated that the propagation distance of the spin current is determined by the spin diffusion length in F, which is typically in a range from a few to 10nm~\cite{Kimura2006,Kimura2007}. So the dissipative current here can be ignored safely and only the supercurrent left. If the bias voltages are applied on the HM electrodes, the spin-dependent quasiparticle distribution can be generated in N regions. It is well known that, in HM electrodes, electronic bands exhibit insulating behavior for one spin direction and metallic behavior for the other. This means that the electrons in the spin-up (spin-down) subband can flow only to or from the top (bottom) electrode. Consequently, the dissipative current flowing through the up and down HM electrodes is non-existent. Besides, the variance of the quasiparticle distribution along $z$-direction is small and can be negligible. Finally, we neglect energy relaxation and spin relaxation processes in the two N regions.

  Let us first consider the quasiparticle distribution function in our model. As shown in fig.~\ref{fig.1}(b), by applying the bias voltage, the spin-dependent quasiparticle distribution $f_{\uparrow,\downarrow}=1/[1+exp\{(\varepsilon_k{\mp}eV)/k_BT\}]$ is created in the N region, where $\varepsilon_k=\frac{\hbar^2{k^2}}{2m}-E_F$ is the one electron energy relative to the chemical potential of the S. Then the electrons forming a pair, which is located at the Fermi level $\varepsilon$ in the S, can only enter the N region with different energies $\varepsilon_{\uparrow,\downarrow}=\varepsilon\pm{eV}$, thus conserving the total energy of the pair. As a result, the difference between the spin-up and spin-down electron momenta will be modified ~\cite{BobPRL}. In this case, the direction of $Q$ is along the HM magnetization axis and $Q\propto2eV/\hbar{v_F}$. Therefore, the spin-dependent quasiparticle distribution will make the electrons enter the N region with unequal (in absolute value) momenta. In this process, we will not take into account the inelastic scattering.

  The Josephson current in the S/N/F/N/S junction, with nonequilibrium quasiparticle distribution in the N regions, is calculated by using Blonder-Tinkham-Klapwijk (BTK) approach~\cite{Blo}. For convenience, the lengths of the trilayer are denoted by $L_{N1}$, $L_F$ and $L_{N2}$. The transport direction is along the \emph{y} axis, and the system is assumed to be infinite in the \emph{x-z} plane. The BCS mean-field effective Hamiltonian~\cite{Buz,Gen} is
  \begin{equation}
  \label{Eq1}
  \begin{aligned}
   H_{eff}&=\int{d\vec{r}}\{\sum_{\tilde{\alpha}}\psi^{\dag}_{\tilde{\alpha}}(\vec{r})[H_e-eV_z(\vec{r})(\sigma_{z})_{\tilde{\alpha}\tilde{\alpha}}]\psi_{\tilde{\alpha}}(\vec{r})\\
   &+\frac{1}{2}[\sum_{\tilde{\alpha},\tilde{\beta}}(i\sigma_{y})_{\tilde{\alpha}\tilde{\beta}}\Delta(\vec{r})\psi^{\dag}_{\tilde{\alpha}}(\vec{r})\psi^{\dag}_{\tilde{\beta}}(\vec{r})+H.C.]\\
   &-\sum_{\tilde{\alpha},\tilde{\beta}}\psi^{\dag}_{\tilde{\alpha}}(\vec{r})(\vec{h}\cdot\hat{\sigma})_{\tilde{\alpha}\tilde{\beta}}\psi_{\tilde{\beta}}(\vec{r})\},
  \end{aligned}
  \end{equation}
  where $H_e=-\hbar^2\nabla^{2}/2m-E_F$, $\psi^{\dag}_{\tilde{\alpha}}(\vec{r})$ and $\psi_{\tilde{\alpha}}(\vec{r})$ are creation and annihilation operators with spin $\tilde{\alpha}$. $\hat{\sigma}$ is the Pauli matrice, and $E_F$ is the Fermi energy. The superconducting gap is given by $\Delta(\vec{r})=\Delta(T)[e^{i\phi_{L}}\Theta(-y)+e^{i\phi_{R}}\Theta(y-L)]$ with $L=L_{N1}+L_F+L_{N2}$. Here, $\Delta(T)$ accounts for the temperature-dependent energy gap. It satisfies the BCS relation $\Delta(T)=\Delta_0\tanh(1.74\sqrt{T_c/T-1})$ with $T_c$ the critical temperature of the Ss. $\phi_{L(R)}$ is the phase of the left (right) S and $\Theta(y)$ is the unit step function. The bias voltage applied on the N regions can be described as $V_z(\vec{r})=V_{1}$ ($0<y<L_{N1}$) and $V_2$ ($L_{N1}+L_{F}<y<L$), and the exchange field in the F layer is characterized as $\vec{h}=h(\sin\alpha\hat{e}_{x}+\cos\alpha\hat{e}_z)$ ($L_{N1}<y<L_{N1}+L_{F}$), where $\alpha$ is the polar angle of the magnetization with respect to \emph{z} axis, and $\hat{e}_{x(z)}$ is the unit vector along the \emph{x}(\emph{z}) direction.

  Based on the Bogoliubov transformation $\psi_{\tilde{\alpha}}(\vec{r})=\sum_{n}[u_{n\tilde{\alpha}}(\vec{r})\hat{\gamma}_{n}+v^{\ast}_{n\tilde{\alpha}}(\vec{r})\hat{\gamma}^{\dag}_{n}]$ and the anticommutation relations of the quasiparticle annihilation and creation operators $\hat{\gamma}_{n}$ and $\hat{\gamma}^{\dag}_{n}$, we have the Bogoliubov-de Gennes (BdG) equation~\cite{Buz,Gen}
  \begin{equation}
  \label{Eq2}
  \begin{pmatrix}
  \hat{H}(\vec{r}) & \hat{\Delta}(\vec{r}) \\
	-\hat{\Delta}^{*}(\vec{r}) & -\hat{H}(\vec{r})\\
  \end{pmatrix}
  \begin{pmatrix}
   \hat{u}(\vec{r})  \\
	\hat{v}(\vec{r}) \\
  \end{pmatrix}
   =E
  \begin{pmatrix}
   \hat{u}(\vec{r})  \\
	\hat{v}(\vec{r}) \\
  \end{pmatrix},
  \end{equation}
  where $\hat{H}(\vec{r})=H_{e}\hat{\textbf{1}}-[eV_z(\vec{r})+h_{z}(\vec{r})]\hat{\sigma}_{z}-h_{x}(\vec{r})\hat{\sigma}_{x}$ and $\hat{\Delta}(\vec{r}) =i\hat{\sigma}_{y}\Delta(\vec{r})$. Here $\hat{\textbf{1}}$ is the unity matrix, $\hat{u}(\vec{r})=(u_{\uparrow}(\vec{r})$, $u_{\downarrow}(\vec{r}))^{T}$ and $\hat{v}(\vec{r})=(v_{\uparrow}(\vec{r})$, $v_{\downarrow}(\vec{r}))^{T}$ are two-component wave functions. The BdG equation can be solved for each S lead, each N layer and F layer, respectively. We have four different incoming quasiparticles, electronlike quasiparticles (ELQs) and holelike quasiparticles (HLQs) with spin up and spin down. For an incident spin-up electron in the left S, the wave function is
  \begin{equation}
  \begin{aligned}
  \Psi^{S}_{L}(y)&=[ue^{i\phi_{L}/2},0,0,ve^{-i\phi_{L}/2}]^{T}e^{ik_{e}y} \\
  &+a_1[ve^{i\phi_{L}/2},0,0,ue^{-i\phi_{L}/2}]^{T}e^{ik_{h}y} \\
  &+b_1[ue^{i\phi_{L}/2},0,0,ve^{-i\phi_{L}/2}]^{T}e^{-ik_{e}y} \\
  &+a'_1[0,-ve^{i\phi_{L}/2},ue^{-i\phi_{L}/2},0]^{T}e^{ik_{h}y} \\
  &+b'_1[0,ue^{i\phi_{L}/2},-ve^{-i\phi_{L}/2},0]^{T}e^{-ik_{e}y}.
  \end{aligned}
  \label{Eq3}
  \end{equation}
  In this process, the coefficients $b_{1}$, $b'_{1}$, $a'_{1}$, and $a_{1}$ describe normal reflection, the normal reflection with spin-flip, novel Andreev reflection, and usual Andreev reflection, respectively. Note that the momentum parallel to the interface is conserved in these processes.

  The corresponding wave function in the right S is
  \begin{equation}
  \begin{aligned}
   \Psi^{S}_{R}(y)&=c_1[ue^{i\phi_{R}/2},0,0,ve^{-i\phi_{R}/2}]^{T}e^{ik_{e}y} \\
  &+d_1[ve^{i\phi_{R}/2},0,0,ue^{-i\phi_{R}/2}]^{T}e^{-ik_{h}y} \\
  &+c'_1[0,ue^{i\phi_{R}/2},-ve^{-i\phi_{R}/2},0]^{T}e^{ik_{e}y} \\
  &+d'_1[0,-ve^{i\phi_{R}/2},ue^{-i\phi_{R}/2},0]^{T}e^{-ik_{h}y},
  \end{aligned}
  \label{Eq4}
  \end{equation}
  where $c_1$, $d_1$, $c'_1$, $d'_1$ are the transmission coefficients, corresponding to the reflection processes described above. The coherence factors are defined as usual, $u=\sqrt{(1+\Omega/E)/2}$, $v=\sqrt{(1-\Omega/E)/2}$ and $\Omega=\sqrt{E^2-\Delta^2}$. $k_{e(h)}=\sqrt{2m[E_F+(-)\Omega]/\hbar^2-k^{2}_{\parallel}}$ are the perpendicular components of the wavevectors with $k_{\parallel}$ as the parallel component.

  The wave function in the $N_p$ ($p$=1, 2) layer is given by
  \begin{equation}
  \begin{aligned}
  \Psi^{N}_{p}(y)&=[f_{p1}\cdot{\exp(ik^{e\uparrow}_{Np}y)}+f_{p2}\cdot{\exp(-ik^{e\uparrow}_{Np}y)}]\hat{e}_{1} \\
  &+[f_{p3}\cdot{\exp(ik^{e\downarrow}_{Np}y)}+f_{p4}\cdot{\exp(-ik^{e\downarrow}_{Np}y)}]\hat{e}_{2} \\
  &+[f_{p5}\cdot{\exp(-ik^{h\uparrow}_{Np}y)}+f_{p6}\cdot{\exp(ik^{h\uparrow}_{Np}y)}]\hat{e}_{3} \\
  &+[f_{p7}\cdot{\exp(-ik^{h\downarrow}_{Np}y)}+f_{p8}\cdot{\exp(ik^{h\downarrow}_{Np}y)}]\hat{e}_{4}.
  \end{aligned}
  \label{Eq5}
  \end{equation}
  Here $\hat{e}_{1}=[1,0,0,0]^{T}$, $\hat{e}_{2}=[0,1,0,0]^{T}$, $\hat{e}_{3}=[0,0,1,0]^{T}$, $\hat{e}_{4}=[0,0,0,1]^{T}$ are basis wave functions, and $k^{e(h)\uparrow,\downarrow}_{Np}=\sqrt{2m[E_F+(-)E\pm{eV_{p}}]/\hbar^2-k^{2}_{\parallel}}$ are the perpendicular components of wave vectors for ELQs and HLQs. Furthermore, the wave function of the F layer can be described by transformation matrix~\cite{Jin} as
  \begin{equation}
  \begin{aligned}
  \Psi^{F}(y)&=T\{[g_1\cdot{\exp(ik^{e\uparrow}_{F}y)}+g_2\cdot{\exp(-ik^{e\uparrow}_{F}y)}]\hat{e}_{1} \\
  &+[g_3\cdot{\exp(ik^{e\downarrow}_{F}y)}+g_4\cdot{\exp(-ik^{e\downarrow}_{F}y)}]\hat{e}_{2} \\
  &+[g_5\cdot{\exp(-ik^{h\uparrow}_{F}y)}+g_6\cdot{\exp(ik^{h\uparrow}_{F}y)}]\hat{e}_{3} \\
  &+[g_7\cdot{\exp(-ik^{h\downarrow}_{F}y)}+g_8\cdot{\exp(ik^{h\downarrow}_{F}y)}]\hat{e}_{4}\}.
  \end{aligned}
  \label{Eq6}
  \end{equation}
  Here $k^{e(h)\sigma}_{F}=\sqrt{2m[E_F+(-)E+\rho_{\sigma}h]/\hbar^2-k^{2}_{\parallel}}$ with $\rho_{\uparrow(\downarrow)}=1(-1)$. The transformation matrix is defined as $T=\hat{\textbf{1}}\otimes(\cos\frac{\alpha}{2}\cdot\hat{\textbf{1}}-i\cdot{\sin\frac{\alpha}{2}}\cdot\hat{\sigma}_{y})$. All scattering coefficients can be determined by solving wave functions at the interfaces
  \begin{equation}
  \begin{aligned}
  &\Psi^{S}_L(y_1)=\Psi^{N}_1(y_1),\partial_{y}[\psi^{N}_{1}-\psi^{S}_{L}]|_{y_1}=2k_FZ_1\psi^{S}_{L}(y_1);\\
  &\Psi^{N}_{1}(y_2)=\Psi^{F}(y_2),\partial_{y}[\psi^{F}-\psi^{N}_{1}]|_{y_2}=2k_FZ_2\psi^{N}_{1}(y_2);\\
  &\Psi^{F}(y_3)=\Psi^{N}_2(y_3),\partial_{y}[\psi^{N}_2-\psi^{F}]|_{y_3}=2k_FZ_3\psi^{F}(y_3);\\
  &\Psi^{N}_2(y_4)=\Psi^{S}_{R}(y_4),\partial_{y}[\psi^{S}_{R}-\psi^{N}_{2}]|_{y_4}=2k_FZ_4\psi^{S}_{R}(y_4).
  \end{aligned}
  \label{Eq7}
  \end{equation}
  Here $y_q$=$0$, $L_{N1}$, $L_{N1}+L_F$, $L$ with $q=1, 2, 3, 4$, and $Z_1-Z_4$ are dimensionless parameters describing the magnitude of the interfacial resistances, as well as $k_F=\sqrt{2mE_F}$ is the Fermi wave vector. From the boundary conditions, we obtain a system of linear equations that yield the scattering coefficients. With this coefficients at hand, we can use the finite-temperature Green's function formalism~\cite{Fur,Zhe,Tan} to calculate dc Josephson current,
  \begin{equation}
  \label{Eq9}
  \begin{aligned}
  &I_{e}(\phi)=\frac{k_BTe\Delta}{4\hbar}\sum_{k_{\parallel}}\sum_{\omega_{n}}\frac{k_{e}(\omega_{n})+k_{h}(\omega_{n})}{\Omega_{n}}\cdot \\
  &[\frac{a_1(\omega_{n},\phi)-a_2(\omega_{n},\phi)}{k_{e}}+\frac{a_3(\omega_{n},\phi)-a_4(\omega_{n},\phi)}{k_{h}}],
  \end{aligned}
  \end{equation}
  where $\omega_{n}=\pi{k_B}T(2n+1)$ are the Matsubara frequencies with $n=0, 1, 2$,\ldots and $\Omega_{n}=\sqrt{\omega^{2}_{n}+\Delta^{2}(T)}$. $k_e(\omega_{n})$, $k_h(\omega_{n})$, and $a_j(\omega_{n},\phi)$ with $j=1, 2, 3, 4$ are obtained  from $k_e$, $k_h$, and $a_j$ by analytic continuation $E\rightarrow{i}\omega_{n}$. Then the critical current is derived from $I_c=max_{\phi}\{|I_e(\phi)|\}$.

  \begin{figure}
  \centering
  \includegraphics[width=3.4in]{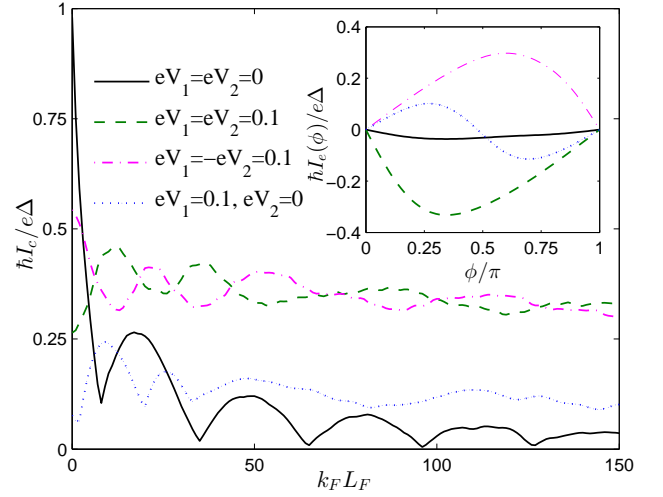} 
  \caption{(Color online) Critical current as a function of the length $L_F$ by applying the different bias voltage. Inset: The Josephson current-phase relation $I(\phi)$ for $k_FL_{N1}=10$ and $k_FL_F=150$.}
  \label{fig.2}
  \end{figure}

  Before showing the specific calculation, we briefly present the choose of the involved parameters. The superconducting gap $\Delta_0$ is set as the unit of energy. The Fermi energy is $E_F=1000\Delta_0$, the interface transparency is $Z_{1-4}=0$, and $T/T_c=0.1$. The exchange field and the polar angle of the F are characterized by fixed value $h/E_F=0.1$ and $\alpha=-\pi/2$, respectively. Interface layers $N_1$ and $N_2$ have the same length $L_{N1}=L_{N2}$.

  In fig.~\ref{fig.2}, we show the dependence of the critical current $I_c$ on the length $L_F$. It is well known, for the bias voltage $eV_1=eV_2=0$, the critical current $I_c$ exhibits oscillations with a period $2\pi\xi_{F}$ and simultaneously decays exponentially on the length scale of $\xi_{F}$~\cite{Buz}. Here, $\xi_{F}$ is the magnetic coherence length. The reason is that only the singlet pairs $\mid\uparrow\downarrow\rangle$$-$$\mid\downarrow\uparrow\rangle$ and triplet pairs $\mid\uparrow\downarrow\rangle$$+$$\mid\downarrow\uparrow\rangle$ exist in the F layer. In contrast, if $eV_1=eV_2=0.1$, we find that the critical current $I_c$ can penetrate a long-range into the F layer, which arises from the propagation of parallel spin triplet pairs. This is because the nonequilibrium spin-dependent quasiparticle distribution in two N regions will provide the Cooper pairs an additional momenta $Q_z$, which is analogous to the momenta induced by the exchange field. Since the direction of $Q_z$ is orthogonal to the magnetization of central F, spin-mixing and spin-flip scattering processes will appear in the N regions. The former process will result in a mixture of the spin singlet pairs $(\mid\uparrow\downarrow\rangle$$-$$\mid\downarrow\uparrow\rangle)_{z}$ and triplet pairs $(\mid\uparrow\downarrow\rangle$$+$$\mid\downarrow\uparrow\rangle)_{z}$. The latter can convert $(\mid\uparrow\downarrow\rangle$$+$$\mid\downarrow\uparrow\rangle)_{z}$ into the parallel spin pairs $\mid\uparrow\uparrow\rangle_{x}$ and $\mid\downarrow\downarrow\rangle_{x}$. This parallel spin pairs will propagate coherently over long distances into the central F layer. It is worth to note that the momenta $Q_z$ induced by the nonequilibrium quasiparticle distribution is not completely equivalent to the one caused by the exchange splitting of the F. The exchange splitting shifts the momenta of the paired electrons, located at the same Fermi energy with opposite spins, from $k_F$ to the new positions $k_{\uparrow}=k_F+Q_z/2$ and $k_{\downarrow}=k_F-Q_z/2$. Then the Cooper pair will obtain an extra total momenta $Q_z$. On the other hand, for the bias voltage, it will shift the Fermi energies from $\varepsilon$ to $\varepsilon\pm{eV}$ for two spin subbands, thus one can obtain the total momenta $Q_z$. In this case, the electrons in the cooper pairs will locate at different Fermi energies and have unequal momenta.

  \begin{figure}
  \centering
  \includegraphics[width=3.5in]{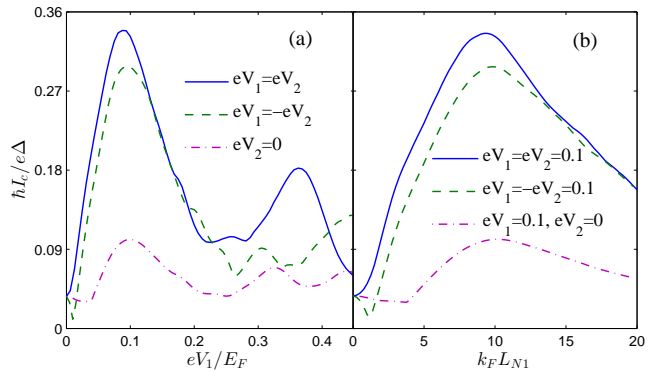} 
  \caption{(Color online) (a) Critical current as a function of the bias voltage $eV$ for $k_FL_{N1}=10$ and $k_FL_F=150$. (b) Critical current as a function of the thickness $k_FL_{N1}$ for $k_FL_F=150$.}
  \label{fig.3}
  \end{figure}

  Furthermore, the Josephson current $I(\phi)$ varying with the directions of the applied voltages is calculated and the results are plotted in the inset of fig.~\ref{fig.2}. If the two applied bias voltages on the two N regions have same directions, $eV_1=eV_2=0.1$, the Josephson current $I(\phi)$ is negative and oscillates with a period $2\pi$. While if the voltages have opposite directions, $eV_1$=$-eV_2$=0.1, the Josephson current $I(\phi)$ will acquire an extra $\pi$ phase and become positive with the same oscillation period. Such sign reversal is similar to the $\pi$ phase shift induced by changing the mutual direction of the ferromagnetic moments in the Josephson junction with ferromagnetic trilayer~\cite{Hou} or the S-F multilayered structures~\cite{AFVolkovPRL}. It means that the direction of the supercurrent can be controlled easily by tuning the direction of applied voltages. As we will discuss in the next paragraph, one also can manipulate the amplitude of the long-range supercurrent through changing the amplitude of the voltages. This method is very promising for artificially controlling Josephson current in the spintronics devices. In addition, if the bias voltage is applied on only one $N_1$ region, the long-range critical current will decrease totally. The reason is that the dominant contribution to the Josephson current in this case stems from the transport of the even number of parallel spin triplet pairs, and this pairs with opposite spin directions have to recombine into singlet Cooper pairs at near the right N region~\cite{Richard}. This behavior is analogous to the presence of only one spin-active region at the SF interface. Interestingly, as shown in the inset of Fig.~\ref{fig.2}, the period of this current now becomes $\pi$ and satisfies the superharmonic current-phase relation $I(\phi)\propto{\sin2\phi}$, so it is the recently discovered long-range superharmonic Josephson current ~\cite{Tri,Richard}.

  Finally, we briefly discuss the dependence of critical current on the voltages at fixed length of N and F, and on the length of the N at fixed voltages. The calculated results are shown in Fig.~\ref{fig.3}. For $eV_1=eV_2$, as plotted in Fig.~\ref{fig.3}(a), the critical current $I_c$ displays an oscillating behavior as a function of $eV$ for $k_FL_{N1}=10$ and $k_FL_F=150$. We attribute this behavior to the oscillations of the spin triplet pairs $\mid\uparrow\downarrow\rangle$$+$$\mid\downarrow\uparrow\rangle_z$ in N layers with period $Q_z$$\cdot$$R$~\cite{Meng}. For the fixed thickness R ($k_FL_{N1}=10$), the total momenta $Q_z$ will vary with the the bias voltage $eV_1$ and $Q_z$$\cdot$$R$ will undergo the $0-\pi$ transitions. As mentioned before, the oscillated $\mid\uparrow\downarrow\rangle$$+$$\mid\downarrow\uparrow\rangle_z$ in N layer can be converted into the $\mid\uparrow\uparrow\rangle_x$ and $\mid\downarrow\downarrow\rangle_x$ in central F layer by the spin-flip scattering. So the dependence of the critical current on the bias voltage has an oscillatory behavior. For $eV_1=-eV_2=0.1$ and $eV_1=0.1$, $eV_2=0$, the critical current has the same behavior. As shown in fig.~\ref{fig.3}(b), for weak voltage, the relationship $Q_z\cdot{R}<\pi$ is always satisfied when $k_FL_{N1}$ increases from 0 to 20. Hence the critical current $I_c$ has only one peak. It is worth noting here that, for $eV_1=eV_2=0.1$, the critical current $I_c$ manifests a monotonous feature, with $k_FL_{N1}$ increasing from 0 to 10. However, for $eV_1=-eV_2=0.1$, $I_c$ exhibits a shallow dip at $k_FL_{N1}=1$. This feature is induced by the phase transition of junction. If one doesn't take absolute value for $I_e(\phi)$ to define the critical current, for $eV_1=eV_2=0.1$, $I_c$ are all negative and decreases monotonously, when $k_FL_{N1}$ changes from 0 to 10. In opposite case, for $eV_1=-eV_2=0.1$, $I_c$ will increase from a negative quantity to a positive one and reach the positive maximum at $k_FL_{N1}$=10, then the non-monotonic behavior mentioned before would transform into monotonic, accompanying with a change of sign. At the same time, for $eV_1=0.1$, $eV_2=0$, $I_c$ have the same the behavior.

  In summary, we have studied the long-range triplet Josephson current in a clean S/N/F/N/S junction. The nonequilibrium spin-dependent quasiparticle distribution in the N regions is produced by applying the bias voltages through two antiparallel HM electrodes. This distribution can provide the Cooper pair in the N regions an extra momenta $Q_z$. Because the direction of $Q_z$ is perpendicular to the magnetization of the central F, this behavior leads to the spin-flip processes in the N regions, which can produce the parallel spin triplet pairs in the central F layer. As a result, when one applies the bias voltages on two N regions, the long-range triplet Josephson current will appear. We have also shown that the direction and amplitude of the Josephson current can be controlled by changing the applied bias voltages. This method is very promising for artificially controlling Josephson current in the practical application of spintronics devices. In addition, if the voltages are applied on only one N region, the recently discovered superharmonic Josephson current can be also produced because of the propagation of the even number of parallel spin triplet pairs in this case.

  This work is supported by the State Key Program for Basic Research of China under Grants No. 2011CB922103 and No. 2010CB923400, and the National Natural Science Foundation of China under Grants No. 11174125 and No. 11074109.

    \end{CJK}
    \end{document}